\documentclass{vldb}
\usepackage{algorithm, algorithmic}
\newcommand{\eat}[1]{}
\pdfoutput=1
\newtheorem{theorem}{Theorem}[section]
\newtheorem{lemma}[theorem]{Lemma}

\newtheorem{definition}[theorem]{Definition}

\newtheorem{example}[theorem]{Example}

\toappear{}
\title{Cooperative Update Exchange in the Youtopia System}
\begin{document}
\numberofauthors{2} 
\author{
\alignauthor
\L ucja Kot\\
       \affaddr{Cornell University}\\
       \affaddr{Ithaca, NY 14853}\\
       \email{lucja@cs.cornell.edu}
\alignauthor
Christoph Koch\\
       \affaddr{Cornell University}\\
       \affaddr{Ithaca, NY 14853}\\
       \email{koch@cs.cornell.edu}
}

\maketitle
\begin{abstract}
Youtopia is a platform for collaborative management and integration of relational data. At the heart of Youtopia is an update exchange abstraction: changes to the data propagate through the system to satisfy user-specified mappings. We present a novel change propagation model that combines a deterministic chase with human intervention. The process is fundamentally cooperative and gives users significant control over how mappings are repaired. An additional advantage of our model is that mapping cycles can be permitted without compromising correctness.

We investigate potential harmful interference between updates in our model; we introduce two appropriate notions of serializability that avoid such interference if enforced. The first is very general and related to classical final-state serializability; the second is more restrictive but highly practical and related to conflict-serializability. We present an algorithm to enforce the latter notion. Our algorithm is an optimistic one, and as such may sometimes require updates to be aborted. We develop techniques for reducing the number of aborts and we test these experimentally.
\end{abstract}

\section{Introduction}

\subsection{Collaborative data integration}

Communities everywhere on the Web want to share, store and query data. Their motivations for data sharing are very diverse -- from entertainment or commercial activity to the desire to collaborate on scientific or artistic projects. The data involved is also varied, running the gamut from unstructured through semistructured to relational. The solutions used for data sharing are frequently custom-built for a concrete scenario; as such, they exhibit significant diversity themselves. To name only a few prominent solutions, Wiki software has proved very successful for community management of unstructured data; scientific portals such as BIRN \cite{Birn} and GEON \cite{Geon} allow scientists to pool their datasets; and an increasingly large number of vertical social networking sites include a topic-specific database that is maintained by the site's members.

While the scenarios mentioned above vary widely in their parameters, they have in common many high-level properties that translate into concrete design desiderata for Collaborative Data Integration (CDI) systems. In the Youtopia project, we are building a system to address these desiderata and enable community data sharing in arbitrary settings. Our initial focus is on relational data; however, the ultimate goal is to include arbitrary data formats and manage the data in its full heterogeneity, as in Dataspaces \cite{HalevyFM06:Principles}.

CDI has three fundamental aspects that distinguish it from other paradigms such as classical data integration. First, a CDI system must enable \emph{best-effort cooperation} among community members with respect to maintenance of the data and metadata. That is, no worthwhile contribution to the repository should be rejected because it is incomplete, as another community member may be able to supply the knowledge required to complete it. This means a CDI system must be equipped to deal with incomplete data and metadata, as well as providing a way for users to complete them at a later time. Next, a CDI solution must \emph{manage disagreement} regarding the data and schema or other metadata. Finally, it must \emph{maximize data utility}.

These three aspects have clear tradeoffs in the extent to which they can be addressed; as such, they define a design space within which we can situate existing solutions and Youtopia. The structure of this design space also clarifies the relationship of CDI to classical data integration; the latter is fundamentally an effort to maintain utility while permitting as much disagreement as possible. CDI builds on this by introducing the added element of best-effort cooperation, familiar from the Web 2.0 model of enabling all users to create their own content on the internet.

\subsection{Youtopia}

\begin{figure}
\begin{center}
\includegraphics[keepaspectratio,width=\columnwidth]{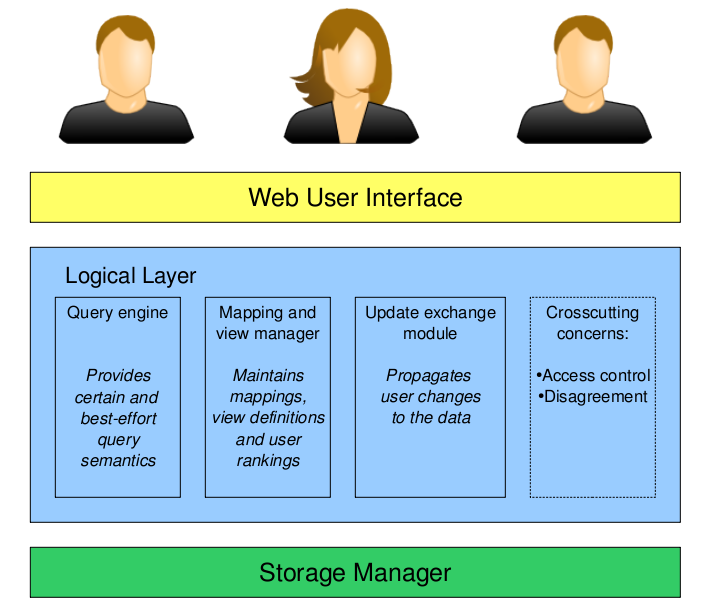}
\end{center}
\vspace{-3ex}
\caption{Architecture of the Youtopia system}\label{fig:architecture}
\vspace{-4ex}
\end{figure}

Youtopia is a system that allows users to add, register, update and maintain relational data in a collaborative fashion. The architecture of Youtopia is presented in Figure \ref{fig:architecture}. The storage manager provides a logical abstraction of the repository. In this abstraction, the repository consists of a set of logical tables or views containing the data; these are tied together by a set of mappings. The mappings are supplied by the users as the repository grows and serve to propagate changes to the data. Thus, at the logical level Youtopia is an \emph{update exchange} system. In this paper, we introduce our update exchange model, which is designed to enable best-effort cooperation as far as possible; in this it differs from previous update exchange work such as Orchestra \cite{GreenKIT07:Update}. 

A small Youtopia repository is shown in Figure \ref{fig:exampledata}. It contains relations with travel and tourist information; the relations are conected by a set of \emph{mappings} or \emph{tuple-generating dependencies (tgds)}. For instance, the tgd $\sigma_3$ ensures that table \texttt{R} contains review information about all available tours of attractions, as explained in the following example.

\begin{example}\label{ex:fwdChase}
Suppose company $\texttt{ABC Tours}$ starts running tours to Niagara Falls and the tuple \texttt{T}(\texttt{Niagara Falls}, \texttt{ABC Tours}) is added. The mapping will cause the new tuple \texttt{R}(\texttt{Niagara Falls}, \texttt{ABC Tours}, $x_3$) to be inserted by the update exchange module. The $x_3$ is a \emph{labelled null} or \emph{variable} which indicates that some review for the tour should exist, but is unknown to the system. The review may subsequently be filled in manually by a user.
\end{example}

This propagation of changes occurs through a process known as the (tgd) \emph{chase} \cite{AhoSU79:Efficient, MaierMS79:Testing, DeutschNR08:Chase} -- a simple mechanism for constraint maintenance in which the corrective operations required are relatively easy to determine and perform.

Tuple-generating dependencies and equivalent constraints such as GLAV mappings \cite{MadhavanH03:Composing} and conjunctive inclusion dependencies \cite{Koch04:Query} are frequently encountered in data integration \cite{GreenKIT07:Update, FaginKMP05:Data, HalevyIST05:Schema, KatsisDP08:Interactive, YanMHFD01:Data}. Their ubiquity points to the fact that they are a very powerful formalism, applicable in a variety of subject domains. On the other hand, it is not always trivial for a user to specify a mapping correctly. However, this problem has been addressed in some existing work \cite{YanMHFD01:Data, SarmaDH08:Bootstrapping} and we are building on these solutions to set up an infrastructure to facilitate  mapping creation. Notably, Youtopia allows users to cooperate and pool their understanding to set up and refine tgds. Mapping creation is also made easier by the presence of subdomain-specific summary views: knowledgeable users can define such views which capture in their schema the essence of the subdomain. Much as portals and topic lists in Wikipedia can guide contributors in the categorization of their articles, such views can guide table owners in the placement of their mappings. 

The subdomain-specific summary views also assist in querying the repository; thus, managing such views is related to collaborative query management \cite{KhoussainovaBGKS09:Case}. Querying in Youtopia involves a mixture of keyword search and structured queries. The query engine is equipped to handle data that is incomplete, inconsistent or both. This is done through the use of multiple query semantics: a \emph{certain} semantics that guarantees correctness while potentially omitting some results, and a \emph{best-effort} semantics that includes all potentially relevant results at the risk of some incorrectness.

Two key crosscutting concerns in Youtopia are disagreement handling and data access control. As data and mappings are added to the repository, disagreement is inevitable. Youtopia provides mechanisms for disagreement resolution by the community, as well as supporting data inconsistency in the event that disagreement is not resolved. Moreover, it provides facilities for community ranking of tables and data. Finally, Youtopia also has a social structure allowing users to establish a network of trusted acquaintances or friends, so that data, mappings, rankings and user-defined views can be shared to a varying extent.

\subsection{Cooperative update exchange}

As we mentioned, the Youtopia update exchange model is designed to maximize potential for best-effort cooperation. Accordingly, there are no centralized constraints on the mappings, such as acyclicity. These restrictions are a bottleneck for extensibility, as users may not be able to determine how a cycle should be broken, particularly if the cycle is complex and includes mappings created by a large number of users. The reason such restrictions are common \cite{GreenKIT07:Update, HalevyIST05:Schema, FaginKMP05:Data, Koch04:Query, CaliGK08:Taming} is that the standard model of update propagation via the chase requires them for termination. We argue that this chase model is not suitable for a CDI setting, as it is stronger than the propagation semantics which are intuitively desired by users when they set up mappings. The infinite chases that can arise are a symptom of this.

In the standard chase with tgds, the system must correct any mapping violations \emph{completely}, \emph{as soon as they occur}, and \emph{without asking for additional information}. In Youtopia, we lift all these requirements, as we believe they are too strong for CDI. Users of a CDI system do not necessarily want mapping violations corrected immediately, particularly if they are not using the relevant part of the repository at the moment. On the other hand, they frequently have domain knowledge that they can supply to the system to aid in violation correction. Based on these observations, we propose a new chase model that is fundamentally cooperative. It turns out that in this model, cycles among mappings no longer cause a problem and can be permitted. 

Of course, human assistance can be slow in coming. To function in a practical setting, the Youtopia system must remain usable as far as possible while violations are waiting for human assistance. This means that neither queries nor new chases can be blocked by an earlier chase that is waiting for human help. As soon as multiple chases are occurring in the system concurrently, it is possible for them to interfere, which affects overall correctness. The challenge is then to allow asynchronous, human-assisted correction of violations without locking down the entire system and while guaranteeing correct semantics for the entire process. 

Our contributions in this paper are as follows. First, we present the Youtopia update model and explain how it combines the classical chase with user intervention. The model introduces the novel concepts of \emph{frontier tuples} and \emph{frontier operations}, where the former represent a point of ambiguity in the chase and the latter provide a means for a human to resolve this ambiguity. The frontier operations are designed to be simple for users who have the appropriate domain knowledge; indeed, they are similar to intuitive data cleaning operations. For example, one of our frontier operations is a \emph{unification}, wherein a user specifies that two tuples refer to the same real-world fact and should be collapsed into one.

Next, we outline the possibilities that arise for interference among multiple Youtopia chases and provide a definition of serializability that can be used to prevent interference. We present a practical strategy for maintaining serializability: we give a concurrency control algorithm framework that can be instantiated in multiple ways. Our general algorithmic approach is optimistic, that is, we allow new updates to set off chases even while older ones are awaiting human assistance. As with any optimistic concurrency control technique, the higher throughput gained comes at a cost: some chases may have to be aborted and restarted if a conflict occurs. We identify mechanisms for reducing the number of such aborts, which are particularly undesirable in the Youtopia setting where a redo may require human intervention. We present experimental results to demonstrate that the number of aborts can indeed be reduced dramatically using our methods. Finally, we discuss issues involved in the implementation of our algorithm in a real system where a scheduler must choose which chases to run and when.

\section{Update exchange in Youtopia}\label{sec:updateModel}

At the logical level, a Youtopia repository is a database, i.e a set of relations containing constants and \emph{labeled nulls} (or variables). The relations are connected by \emph{mappings} or tuple-generating dependencies \cite{FaginKMP05:Data, TaylorI06:Reconciling}. A mapping specifies a desired relationship between the data in the relations which it connects. It has the form $$\Phi(\overline{x}, \overline{y}) \rightarrow \exists \overline{z} \Psi(\overline{x}, \overline{z})$$ where $\Phi$ is a conjunction of relational atoms over the sets of variables and constants $\overline{x}$ and $\overline{y}$, while $\Psi$  is a conjunction of relational atoms over $\overline{x}$ and $\overline{z}$. The free variables are understood to be universally quantified.

\begin{figure}
\includegraphics[keepaspectratio,width=\columnwidth]{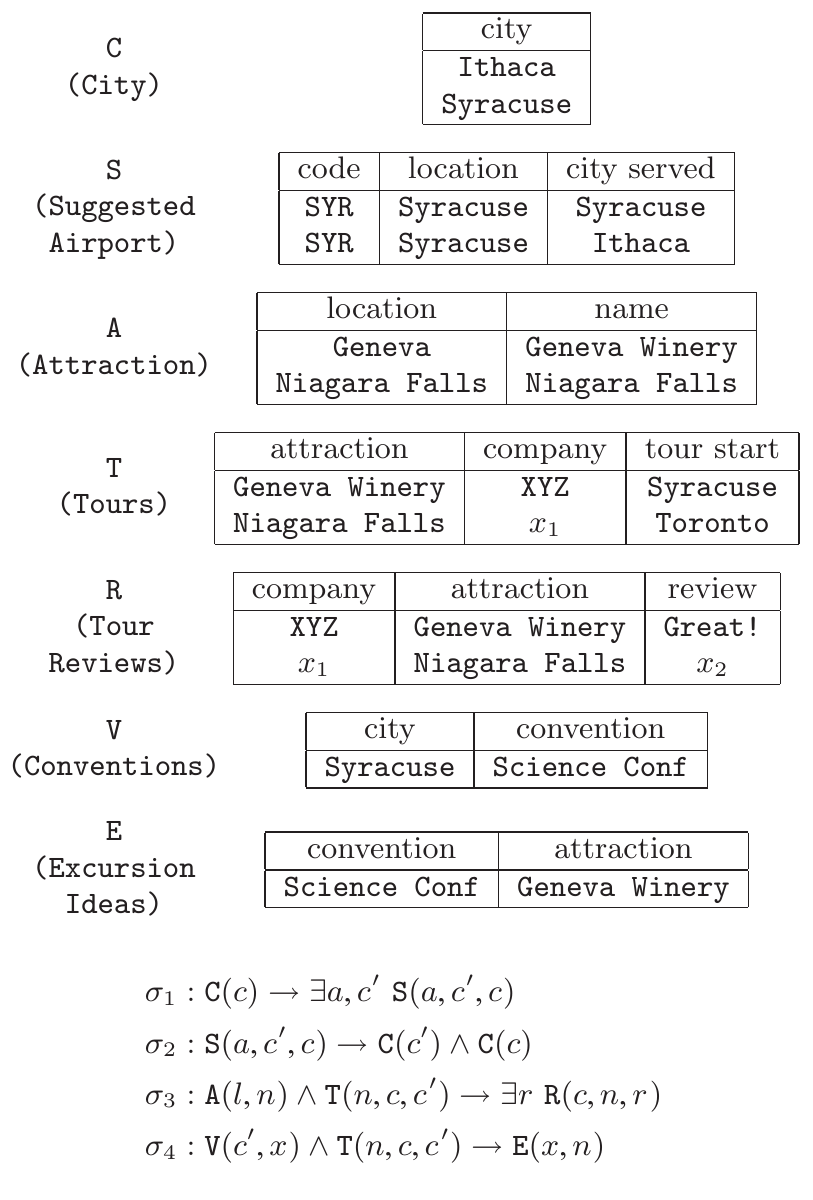}
\vspace{-7ex}
\caption{Example Youtopia repository. $\sigma_1$ states that every city has a recommended airport. Under $\sigma_2$, every airport is located in a city and serves a city. $\sigma_3$ ensures that whenever a company offers tours of an attraction, the tour is reviewed. Because of $\sigma_4$, convention attendees can receive recommendations for day trips based on the convention venue and available tours.}\label{fig:exampledata}
\vspace{-2ex}
\end{figure}

Figure \ref{fig:exampledata} shows an example of a small Youtopia repository and mappings. Mappings in Youtopia may connect arbitrary relations, they may include features such as self-joins, and -- most importantly -- they may form cycles over the set of relations. For example, $\sigma_1$ and $\sigma_2$ form a cycle involving the relations \texttt{C} and \texttt{S}. 

When a tuple is inserted, deleted or modified by a user, some of the mappings between relations may no longer be satisfied, that is, violations may occur.

\begin{definition}[Violation]\label{def:violation}
Let $\sigma$ be a tgd that is not satisfied by database $D$. Use $\bar{x}$ to denote the free variables of $\sigma$, $\bar{y}$ the existentially quantified ones, $\sigma_l(\bar{x})$ the LHS subformula of $\sigma$ and $\sigma_r(\bar{x}, \bar{y})$ the RHS. A \emph{violation} is an assignment $\overline{a}$ of values from the database domain to $\overline{x}$ such that $D \vDash \sigma_l(\overline{x})[\overline{x} \mapsto \overline{a}]$ but $D \nvDash \sigma_r(\overline{x}, \overline{y})[\overline{x} \mapsto \overline{a}]$.
\end{definition}

\begin{definition}[Witness]\label{def:witness}
Every violation of a mapping $\sigma$ is associated with a \emph{witness} - this is the set of tuples from the database that match the LHS of $\sigma$, but that do not have a corresponding set of tuples to match the RHS. 
\end{definition}

Youtopia uses a variant of the standard chase procedure \cite{AhoSU79:Efficient, MaierMS79:Testing} to correct any new mapping violations. In this paper, we restrict ourselves to three kinds of user operations that may initiate a chase: tuple insertion, tuple deletion, and \emph{null-replacement} -- replacement of all occurrences of a labeled null with the same constant value.

If a violation is caused by a tuple insertion or by tuple change due to null-replacement, this must be because the new version of the tuple is part of this violation's witness. We call this type of violation an LHS-violation. It is clear why insertions only cause LHS-violations, but perhaps less so for null-replacements. In particular, this is \emph{not} generally true for arbitrary tuple modifications. However, in null-replacements, all instances of a labeled null are changed simultaneously and consistently, and this ensures that only LHS-violations are possible. For example, if $x_1$ were to be replaced by \texttt{ABC Tours}, a violation of $\sigma_3$ would not occur because the change would happen in both \texttt{T} and \texttt{R}.

On the other hand, suppose a mapping is violated because of a tuple deletion. This must be because the deleted tuple used to be reflected on the RHS of some assignment of values, but now that the tuple is gone, no other matching tuple to the same RHS atom can be found. We call this type of violation an RHS-violation.

\subsection{Violation repair}

The Youtopia chase has two variants - \emph{forward} and \emph{backward}. The forward chase corrects violations by supplying the missing RHS tuples for the violation witness, as in Example \ref{ex:fwdChase}. The backward chase corrects violations by removing at least one of the witness tuples. 

\begin{example}\label{ex:backwardChase}
Suppose the tuple \texttt{R(Geneva Winery, XYZ Tours, Great!)} is deleted from our example database. A backward chase corrects the violation of $\sigma_3$ by deleting either \texttt{A(Geneva, Geneva Winery)} or \texttt{T(Geneva Winery, XYZ Tours)}.
\end{example}

In Youtopia, the choice of chase variant is dictated by the type of the violation. LHS-violations are always repaired by a forward chase, RHS-violations by a backward chase. Our primary motivation in making this design choice is the assumption that the user's initial operation accurately reflects their intent with respect to the relation involved. As such, this operation should not be undone by the violation correction process; for example, an insertion that created a LHS-violation should not immediately be undone by a backward chase aiming to correct the violation.

In a real-world setting, users will desire additional functionality on top of our violation repair semantics, particularly with respect to the handling of deletions. Our handling of RHS-violations effectively cascades all deletions. This may be dangerous from an access control standpoint; the system must check whether a deletion may cascade into a table where it would cause a permissions violation. In such a case, the original deletion should be rejected by the system. If the user still wishes the rejected deletion to occur on the original table, they are in effect requesting an exception to the mapping; developing support for such exceptions is ongoing work. Moreover, the user interface must make it easy for the user to determine whether they really intend to perform a deletion. In Example \ref{ex:backwardChase}, the user's intent may have been just to remove the review rather than delete the entire tuple; if so, they should have replaced \texttt{Great!} by a fresh labeled null instead. This would have exactly the same result as performing the original deletion and repairing it with a \emph{forward} chase (i.e., through regeneration). However, a regeneration implementation obscures the user's intent, whereas the direct replacement of a value by a null makes it clear.

We note that while the names of our two chase methods may suggest a similarity to the \emph{chase and backchase (C \& B)} technique \cite{DeutschPT99:Physical}, there is not a close relationship between C \& B and our work. In C \& B, the two chases proceed in distinct phases while ours are interleaved; moreover, C \& B is a mechanism for query optimization on a given database, while our chases serve to propagate changes to the data.

\subsection{The Youtopia forward chase}

Superficially, the forward chase in Youtopia is very similar to the standard tgd chase \cite{FaginKMP05:Data}. A witness is identified: matching RHS tuples are generated, and -- in the example we have seen so far -- inserted into the database. However, if the insertion were always carried out, it would sometimes be possible to generate an infinite cascade of inserts. Suppose JFK airport is added as a suggested access airport for Ithaca. The tuple \texttt{S(JFK, NYC, Ithaca)} is added to the database. To satisfy mapping $\sigma_2$, we need to add tuple \texttt{C(NYC)}. This in turn causes a violation of $\sigma_1$, which can be fixed by inserting \texttt{S($x_3$, $x_4$, NYC)}. This new insert causes a violation of $\sigma_2$, requiring the insertion of \texttt{C}($x_4$), and so on. Such cyclical firing of rules is a well-known problem and the main reason behind mapping acyclicity restrictions \cite{Koch04:Query, KatsisDP08:Interactive, GreenKIT07:Update, FaginKMP05:Data}. 

However, tuple insertion is not the only way to repair a violation by supplying a matching RHS to the witness. Sometimes it is possible to provide a matching RHS by \emph{unifying} some labeled nulls with other values in the database. Indeed, if a knowledgeable human were observing the infinite sequence of inserts mentioned previously, they would very likely step in and short-circuit the process. For example, they might supply the additional information that the suggested airport for NYC is itself in NYC. This is equivalent to indicating that the two tuples  \texttt{C($x_4$)} and \texttt{C(NYC)} are referring to the same fact and should be collapsed. 

(Lest the user's intervention should appear obvious and our entire example contrived, we observe that there is seldom a perfect correspondence between the airport(s) in a city and those recommended for access to the city. A small city is often best served by airports in neighboring locations. Further, an airport in city A may be recommended for travel to city B but not to A itself -- perhaps it is a minor airport with difficult access to downtown A but within short driving distance of B. Thus, users may legitimately want to keep their mappings as flexible as $\sigma_1$ and $\sigma_2$.)

The Youtopia forward chase model is a formalization of the above intuition. A forward chase starts out in the traditional way: we identify violations and their witnesses, generate new tuples and insert them. This chase's execution sequence can be represented as a tree, with inserted tuples as nodes and direct causality relationships as edges. If on any path the system detects nondeterminism such as was present in our example with respect to the tuple \texttt{C($x_4$)}, the chase stops along that path and awaits human intervention. Our notion of nondeterminism is based on the concept of the \emph{more specific than} relation on tuples.

\begin{definition}[Specificity Relation] \label{def:specific}
A tuple $t = (a_1, a_2, \cdots a_k)$ is \emph{more specific than} a tuple $t' = (a'_1, a'_2, \cdots a'_k)$ if the mapping $f$ defined as $f(a'_i) = a_i$ is a function and $f$ is the identity on constants (i.e. values that are not labeled nulls)
\end{definition}

We say that nondeterminism occurs on a chase path if a tuple $t$ belonging to relation $R$ is generated by the chase, but $R$ already contains a tuple $t'$ which is more specific than $t$. In this case, it is possible that $t$ is intended to represent the same fact as $t'$, and $t$ should be set aside for human inspection to determine whether this is the case. In the above example, the tuples \texttt{C(NYC)} and  \texttt{S($x_3$, $x_4$, NYC)} would be inserted by the chase, as no tuples more specific than these exist. On the other hand, the tuple \texttt{C($x_4$)} would not be inserted, since more specific tuples do exist. In this way, the chase would stop even though a cycle of mappings exists. Indeed, it is always the case that such a forward chase must stop sooner or later.

\begin{lemma}\label{lemma:fwdChaseStops}
For any forward chase using the above algorithm, computation will stop along all paths in the chase tree after finitely many steps, unless the chase terminates before such a point is reached.
\end{lemma}

Once a chase has stopped, it is time for a human user to step in and assist the process. The user has access to all the tuples which were generated but not inserted into the database; we refer to those as \emph{positive frontier tuples}. Faced with a frontier tuple $t$, a user may perform one of two \emph{frontier operations}:

\begin{itemize}
	\item \emph{expand} $t$, that is, insert $t$ into the database.
	\item \emph{unify} $t$, that is, choose another tuple $t'$ in the same relation as $t$ which is more specific than $t$ and perform variable unification between any labeled nulls in $t$ and $t'$. $t$ disappears after such an operation.
\end{itemize}

Given a suitable interface that provides meaningful provenance information for the frontier tuples, such frontier operations should be quite feasible for a knowledgeable human to perform. The user is simply presented with a frontier tuple and asked: ``Is this a new tuple, or can you match it to a tuple already in the relation?''. If they answer yes to the first option, they are requesting expansion, and otherwise the matching tuple they indicate supplies the necessary unification information.

The unification operation may cause changes to multiple tuples in the system if they contained one of the labeled nulls which disappeared in the unification. These changes may themselves cause further mapping violations. Expansion may also generate new violations due to the insertion of $t$. However, in both cases the new violations are guaranteed to be LHS-violations, so the chase can simply add them to its violation queue for future correction.

A special case for frontier operations concerns tgds with multiple atoms on the right-hand side. The firing of such a tgd in a forward chase generates multiple frontier tuples that may share some labeled nulls. On such tuple sets, the frontier operations work as expected given that the shared labeled nulls must be treated consistently.

After a frontier operation, the system may be able to carry out further deterministic chase steps; if it can, it does so until it terminates or once again reaches a point where it must stop on all paths. At this point it asks for user assistance again and the process repeats. A chase execution thus consists of a sequence of \emph{deterministic strata} separated by periods of blocking and waiting for frontier operations.

The full forward chase execution model is presented in Algorithm \ref{alg:fwdChase}.

\begin{algorithm}
\caption{The forward chase}
\label{alg:fwdChase}
\begin{algorithmic}
\STATE \texttt{writeSet} := initial user operation
\STATE \texttt{violQueue} := $\emptyset$
\REPEAT
\STATE \COMMENT{Begin deterministic stratum}
\WHILE{\texttt{!writeSet.isEmpty}}
\STATE perform writes in \texttt{writeSet}
\STATE \texttt{writeSet} := $\emptyset$
\STATE \texttt{violQueue.append}(violations just created)
\STATE choose \texttt{v $\in$ violQueue}
\IF {\texttt{v} can be corrected without ambiguity}
\STATE \texttt{writeSet} := set of corrective writes for \texttt{v}
\STATE remove \texttt{v} from \texttt{violQueue}
\ELSE
\STATE generate and store frontier tuples for \texttt{v}
\STATE make nonblocking request to user for frontier op
\ENDIF
\ENDWHILE
\STATE \COMMENT{End deterministic stratum}
\STATE block while no frontier operations performed
\STATE \texttt{writeSet} := result of first frontier operation received
\STATE remove corresponding \texttt{v} from \texttt{violQueue}
\UNTIL{\texttt{writeSet.isEmpty}}
\end{algorithmic}
\end{algorithm}

\subsection{The Youtopia backward chase}

Unlike a forward chase, a backward chase must eventually terminate, as it cannot delete more tuples than exist in the database initially. However, backward chases come with their own flavor of nondeterminism which also requires human assistance to resolve. In Example \ref{ex:backwardChase} it was sufficient for one of the two tuples in question to be deleted for $\sigma_3$ to be satisfied again; the Youtopia system recognizes this, but does not make a decision, deferring it instead to a user.

Like the forward chase, the backward chase progresses deterministically as far as it can; when it encounters a situation like the above, with a set of tuples any of which may be deleted, it marks all these tuples as \emph{negative frontier tuples} and requests user assistance. Faced with such a set of negative frontier tuples, a user may perform the \emph{negative frontier operation} of deleting any subset of the tuples.

Once again, performing this frontier operation requires no technical knowledge from the user. They are simply presented with a set of tuples and requested to select the subset which is to be deleted based on their domain knowledge.

We remark that the deletion frontier operation is in some sense the counterpart of the expansion operation on the positive frontier; both cause the frontier to advance further. It is also possible to formulate a negative frontier operation that would be the counterpart of unification; this would be a \emph{reconfirmation} operation, where a user specifies for some proper subset of a set of negative frontier tuples that the subset is \emph{not} to be deleted. Determining whether such an operation would be useful in a CDI system and, if so, implementing support for it, is future work.

\subsection{Youtopia updates}

As should be clear from the chase descriptions, the effects of a single user operation can propagate in a Youtopia system for many steps. We define a \emph{Youtopia update} to refer to all these consequences together.

\begin{definition}[Update]
An \emph{update} is the complete sequence of database modification operations induced by a single initial tuple insertion, deletion or null-replacement. This includes any frontier operations taken by a user on frontier tuples generated by the update. An update may be nonterminating. An update is \emph{positive} if the initial operation was an insertion or null-completion, and \emph{negative} if the operation was a deletion.
\end{definition}

The fundamental importance of human intervention to our chase models might appear to make them impractical in the real world. However, Youtopia is carefully designed to minimize the system usability impact of waiting for human input. New updates are allowed to begin and proceed while older ones are blocking; this raises potential interference issues, of course, and the remainder of our paper presents our solution for addressing those. 

Moreover, the repository is available for querying at all those times when no chase is executing a deterministic stratum (that is, almost all the time). Of course, the mappings are not necessarily satisfied in the system, which might pose a problem for queries. However, if the system is in a state where all ongoing updates are waiting for frontier operations, it actually satisfies the mappings under a \emph{weaker} semantics. This semantics is based on epistemic logic and similar to the semantics for peer-to-peer data integration given in  \cite{CalvaneseGLR04:Logical} and \cite{FranconiKLS03:Robust}. The positive frontier tuples are not certain with respect to their presence in the appropriate tables; all tuples which are certain are present. A symmetric statement can be made about the negative frontier tuples. Developing suitable query semantics for Youtopia under these epistemic mapping semantics is ongoing work.

\section{Update serializability}

We now introduce a definition of serializability for Youtopia updates. We assume that the initial state of the system consists of a finite database $D$ that satisfies all mappings. There is a finite set of updates  $U$ to be performed, and a total order on the priority of these updates is given. For example, the updates can be ordered by their timestamp or explicitly by the user; we do not assume anything about the ordering beyond the fact that it is total, and updates with a lower number in the order have higher priority.

Defining serializability in Youtopia is complicated by the fact that updates may never terminate. Therefore, serializability cannot be a property of complete excecution traces or schedules, as there are legal execution scenarios which never yield complete traces. Instead, we define it as a property of a finite prefix of an execution schedule. Our definition is based on the classical concurrency theory concept of \emph{final-state serializability} \cite{WeikumV02:Transactional}; however, it explicitly takes into account the semantics of the chase, in that we only consider (prefixes of) schedules which correspond to valid chase executions.

We motivate our serializability definition with an example.

\begin{example}
Consider the database in Figure \ref{fig:exampledata} and suppose the two following real-world events occur. First, company XYZ discontinues tours to the Geneva Winery; the owner of the \texttt{R} table learns about it and removes \texttt{R(XYZ, Geneva Winery, Great!)}. Second, a new conference, \texttt{Math Conf}, is scheduled to take place in Syracuse and the tuple \texttt{V(Syracuse, Math Conf)} is inserted. Both these operations set off updates, numbered $1$ and $2$ respectively. Suppose the following interleaving of events occurs:
\end{example}

\begin{enumerate}
\item $1$ deletes \texttt{R(XYZ, Geneva Winery, Great!)}, causing a violation of $\sigma_3$. The violation cannot be corrected deterministically, so the system marks \texttt{A(Geneva, Geneva Winery)} and \texttt{T(Geneva Winery, XYZ, Syracuse)} as deletion candidates and requests a frontier operation.
\item $2$ inserts \texttt{V(Syracuse, Math Conf)}. This causes a violation of $\sigma_4$ which can be corrected deterministically.
\item $2$ inserts \texttt{E(Math Conf, Geneva Winery)}.
\item $1$ receives a frontier operation directing it to delete \texttt{T(Geneva Winery, XYZ, Syracuse)}.
\end{enumerate}

The resulting database is not the same as it would have been if $2$ had executed serially after $1$. Moreover, we can diagnose the problem before the schedule completes -- it is already apparent by step $3$. Update $2$ has decided prematurely that it should insert a tuple into $E$, even though one of the tuples in the witness to the violation of $\sigma_4$ was a deletion candidate. 

To ensure serializability, we must rule out all execution schedules which perform such premature operations. That is, we must exclude schedule prefixes for which there is a possible future scenario where
\begin{itemize}
\item all running updates terminate, but
\item the resulting schedule is not final-state serializable, and moreover the problem can be traced back to the current schedule prefix.
\end{itemize}

\subsection{Chase model and schedules}

We now work towards making the above intuition more formal. Our first step is to model a Youtopia update in a way that abstracts away the fact that human input is involved in its execution; this is not relevant for our purposes. We only require a way to capture an update's interaction with the database. Therefore, we move from the model in Algorithm \ref{alg:fwdChase} to a model where a chase is a sequence of \emph{steps}. Each step may or may not include a blocking operation where input is requested from a user. Algorithm \ref{alg:chaseStep} gives our model for execution of a single chase step. Here and in the remainder of the paper, the term \emph{chase} refers to any Youtopia chase, forward or backward, unless explicitly specified otherwise.

\begin{algorithm}
\caption{A chase step in Youtopia}
\label{alg:chaseStep}
\begin{algorithmic}
\STATE perform a set $W$ of write operations 
\FORALL{mappings $\sigma$ potentially violated by a write in $W$}
\STATE \texttt{$V_{\sigma}$} := set of new violations of $\sigma$
\STATE \texttt{violQueue.append($V_\sigma$)}
\ENDFOR
\IF {\texttt{violQueue} contains a deterministically repairable $v$}
\STATE \COMMENT{step belongs to deterministic stratum}
\STATE \texttt{nextViol} := $v$
\ELSE 
\STATE choose \texttt{nextViol} from \texttt{violQueue}
\ENDIF
\STATE generate a set $W'$ of corrective writes to repair \texttt{nextViol}
\STATE \COMMENT{generation may include blocking request for user input}
\IF {$W'$ also repairs some $v' \in$ \texttt{violQueue}}
\STATE \texttt{violQueue.remove($v'$)}
\ENDIF
\end{algorithmic}
\end{algorithm}

Algorithm \ref{alg:chaseStep} exposes the write and read operations a chase step performs as it is executed. The write operations occur first, at the beginning of every step; each write operation is either a tuple insertion, deletion, or an update which is part of a global replacement of a variable by a new value. Subsequently, the update step may perform reads for two reasons: to determine what new violations were caused by the writes, and to correct the violation \texttt{nextViol}. This is explained in detail in Section \ref{subsec:safety}; for now, note as an example that correcting a LHS-violation requires reading the database to determine whether it contains any tuples more specific than the frontier tuple. 

The presentation of Algorithm \ref{alg:chaseStep} reflects our first simplifying assumption -- we assume all the writes are performed before the reads begin. This is reasonable: in the chase, the data read in a step is virtually certain to have been modified by the writes that were performed at the start of the same step. Delaying the reads until the writes have completed is therefore necessary for correctness.

An update's execution on the database can thus be completely described by a sequence of alternating sets of write and read operations. We call such a sequence a (single-update) \emph{schedule}. A schedule can be finite or infinite.  We use $\tau$ to denote schedules; if $D$ is a database, we denote by $\tau(D)$ the database that results when $\tau$ is executed over $D$.

The notion of a schedule extends to a multi-update setting; we can assume that each operation in the multi-version case is tagged with the priority number of the corresponding update. In the multi-update setting, arbitrary interleavings may occur between operation sets belonging to different updates; we refer to a \emph{single-update projection} of a schedule to mean the ordered portion of the schedule consisting only of the steps performed by a specific single update.

\begin{definition}[Validity]
A schedule is \emph{valid} with respect to a database $D$ iff it represents an execution of the updates on $D$ that is syntactially and semantically correct with respect to all rules of the chase.
\end{definition}

\begin{definition}[Terminating extension]
A \emph{terminating extension} $\tau'$ to a schedule $\tau$ for update $u$ with respect to database $D$ is a valid schedule consisting of operations belonging to $u$, such that when $\tau \cdot \tau'$ completes its execution on $D$, $u$ terminates.
\end{definition}

\begin{definition}[Serialization]
The \emph{serialization} of a schedule $\tau$, denoted $S(\tau)$, is a schedule obtained by sorting the operations in $\tau$ on the update number, in ascending order (but retaining the ordering between operations of a single update).
\end{definition}

\subsection{Serializability}

As indicated before, we define serializability on schedule prefixes. According to our definition, the serializable schedule prefixes are exactly those which do not ``already'' contain a mistake that would make serialization impossible in some possible future where all updates terminate. To ensure the mistake can be traced back to the current schedule, the definition's scope is restricted to possible futures in which all updates execute and terminate serially. 

\begin{definition}[Final-state Serializability]\label{def:serializability}~
Assume $k$ updates are running in the system. A Youtopia schedule $\tau$ is \emph{final-state serializable} if there exists an ordering of the updates such that for every extension $\tau' = \tau_1 \cdot \tau_2 \cdots \tau_k$, where $\tau_i$ is a terminating extension for update $i$ with respect to database $\tau \cdot \tau_1 \cdot \tau_2 \cdots \tau_{i-1}(D)$, we have that $S(\tau \cdot \tau')$ is valid and $S(\tau \cdot \tau')(D) = \tau \cdot \tau'(D)$. The schedule is serializable \emph{with respect to a given ordering} if the above is true for the specific ordering desired.
\end{definition}

In the case where $\tau$ already is a terminating schedule, the set of possible $\tau'$ is empty, so the condition reduces to testing that $S(\tau)(D) = \tau(D)$, which is exactly traditional final-state serializability.

\section{Concurrency control}

Achieving serializability clearly requires restricting interleavings of operation sets that belong to different updates. Definition \ref{def:serializability} is very general and open to enforcement by a wide range of solutions. In standard concurrency control settings, final-state serializability is usually enforced indirectly by maintaining stronger properties such as conflict serializability \cite{WeikumV02:Transactional} which are also easier to check in practice. Our approach is similar; we begin with restricting our system model and then present our algorithmic framework for enforcing properties that imply serializability.

Our solution draws on existing work on multiversion concurrency control protocols, notably the abort-based MVTO and locking-based MV2PL \cite{WeikumV02:Transactional}. It also makes use of some ideas from predicate and precision locking \cite{EswaranGLT76:Notions, JordanBB81:Precision}. These are well-known protocols and algorithms; our contribution in this paper is to show how they may be adapted to produce a complete solution for enforcing serializability in our new update exchange model.

\subsection{Assumptions and system model}\label{subsec:multiversionIntro}

Our first simplifying assumption is to place an interleaving restriction. We assume that interleavings are only permitted at the chase step granularity. That is, if an update has started a chase step and is performing writes, it can finish the writes and perform all the reads it requires in that step before any other update may proceed with any operation. This restriction need not actually be enforced to the letter physically, as long as the system maintains the illusion that it holds at the logical level. We consequently move to a model where we introduce a \emph{scheduler} component that controls the execution of the updates in the system (Algorithm \ref{alg:chaseScheduler}. The scheduler may permit interleavings at the level of individual steps or allow updates to run an entire deterministic stratum before regaining control; it may also use a wide variety of scheduling policies in choosing which update gets to run.

\begin{algorithm}
\caption{Chase scheduler}
\label{alg:chaseScheduler}
\begin{algorithmic}
\STATE \texttt{writeQueue} := initial writes of each update to be run
\WHILE{\texttt{!writeQueue.isEmpty}}
\STATE choose a set of writes $W$ to schedule from \texttt{writeQueue}
\STATE run one or more chase steps for this update (Alg. \ref{alg:chaseStep})
\STATE enqueue new writes $W'$ to \texttt{writeQueue}
\ENDWHILE
\end{algorithmic}
\end{algorithm}

This interleaving restriction makes it possible to formulate a practical, simple concurrency control algorithm based on a notion of conflict-serializability. Intuitively, it is clear that we need two properties to hold. First, given an update $u$ with number $j$, $u$'s writes must not ``pollute'' the future reads of lower-numbered updates. Second, $u$ should not read data items which will (or may) still be written by lower-numbered updates. We ensure the first property by using tuple \emph{versions} as is done in MVCC algorithms, and the second by defining and enforcing \emph{chase step safety}. 

Our use of versioning is relatively standard for multiversion algorithms \cite{WeikumV02:Transactional}. For each tuple, the database maintains multiple versions. A version is created whenever the tuple is inserted, modified through a null-completion, or deleted. If an update operates on a tuple twice, two versions are created, the second with a higher number. At any time, for update $u$ with number $j$ the visible version of a tuple $t$ is the one with the largest number among those created by any update with (priority) number less than or equal to $j$.

The second half of enforcing serializability involves preventing premature reads. A precise definition requires formalizing the reads performed by a chase step.

\subsection{Read queries}

When an update $u$ executes a chase step, the set of tuples read depends on the contents of the database as well as the information known before the step begins (i.e., the writes in $W$ and the current violation queue). The same write, performed on different databases, will generate different new violations. Because of this, we represent the set of tuples which are read in $C$ intensionally; we identify it with the answers to a set of \emph{read queries} that are performed by $u$ while executing this step. 

An update performs read queries for two reasons: to identify new violations caused by a write, and to obtain information required for violation correction. In the former case, the query to be asked for each mapping $\sigma$ has the form:

\begin{verbatim}
SELECT * FROM (LHS query)
WHERE NOT EXISTS (SELECT * FROM (RHS query))
\end{verbatim}

\texttt{LHS query} and \texttt{RHS query} are conjunctive queries whose structure is dictated by $\sigma$ and bindings by the newly written tuple. We refer to this type of query as a \emph{violation query}. 

\begin{example}\label{ex:violationquery}
Returning to our database in Figure \ref{fig:exampledata}, if the tuple \texttt{R(XYZ, Geneva Winery, Great!)} is deleted, the query to discover violations of $\sigma_3$ is shown below. This query returns all the pairs of \texttt{A} and \texttt{T} tuples which have been affected by the deletion with respect to satisfying $\sigma_3$. 

\begin{verbatim}
SELECT * FROM A, T 
WHERE A.name = T.attraction AND
AND A.name = 'Geneva Winery' AND T.company = 'XYZ'
AND NOT EXISTS (SELECT * FROM R
                WHERE R.company = T.company
                AND R.attraction = A.name)
\end{verbatim}
\end{example}

An update may also perform queries in order to determine how to correct a violation, with or without human help. In the case of RHS-violations, no further reads are performed - the system or a human chooses a tuple to delete from one of the already-read witness tuples. For LHS-violations, however, additional reads may be performed because of the possibility of correction through unification. Given a set of frontier tuples for a violation, the system must perform the following queries for each frontier tuple $t$ belonging to relation $R$:

\begin{itemize}
\vspace{-0.7ex}
\item find any $t' \in R$ more specific than $t$
\vspace{-1.0ex}
\item if such $t'$ are found, for all labeled nulls $x$ in $t$ which were $\emph{not}$ freshly generated when $t$ was created, find all other tuples in the database containing $x$. If unification is chosen, all such tuples must be updated.
\vspace{-0.7ex}
\end{itemize}

We call these types of queries \emph{correction queries}.

\subsection{Safety and conflict-serializability}\label{subsec:safety}

With the notion of read queries in place, we can now define Youtopia conflict-serializability. Our definitions assume that interleaving only occurs at chase step granularity and that versioning is in place so that the only writes visible to an update are those of lower-numbered updates and its own.

\begin{definition}[Safety of a step]
Chase step $C$ is \emph{safe} to perform for update $u$ with number $j$ on database $D$ iff for all $\tau' = \tau_1 \cdot \tau_2 \cdots \tau_{j-1}$, where each $\tau_i$ is a terminating extension for update number $i$ with respect to database $\tau \cdot \tau_1 \cdot \tau_2 \cdots \tau_{i-1}(D)$, we have that for all read queries $q$ that $u$ may perform in step $C$,
\begin{equation*}
q(D) = q(\tau \cdot \tau'(D)).
\end{equation*}
\end{definition}

Safety is closely related to precision locking \cite{JordanBB81:Precision}, since it brings together intensionally specified reads and extensionally specified writes. However, our definition is explicitly tied to the semantics of the Youtopia chase.

\begin{definition}[Conflict serializability]
A schedule is \emph{conflict serializable} iff no chase step is performed by an update until this step is safe.
\end{definition}

\begin{theorem}\label{thm:safetyGivesSerializability}
Any Youtopia schedule that is conflict serializable is final-state serializable (Definition \ref{def:serializability}). The order of serializability is the same as that used when determining safety.
\end{theorem}

\section{Algorithm}

Conflict serializability in Youtopia can in principle be enforced in two ways. We can take a blocking approach, as in MV2PL \cite{WeikumV02:Transactional}, and prevent chase steps from proceeding until it is safe to do so. Alternatively, we can take an optimistic approach where we are more permissive about chase step scheduling, but take the risk that conflicts may occur and may need to be resolved through aborts -- as in MVTO. 

The pure blocking approach is unlikely to be applicable in practice. First, the computational complexity of deciding whether a chase step is safe is currently unknown, and we suspect that it is high. Second, there is always a real possibility that an update that is given permission to proceed by a blocking scheduler could itself block for a long time while awaiting frontier operations. This locks down the entire system and may last indefinitely. Thus, an optimistic approach is necessary, to allow updates to proceed even while others are waiting for user input. Of course, aborts are always undesirable, and particularly so in Youtopia, where an update redo may require repeated frontier operations from users. Thus, the number of aborts must be minimized.

The core of our optimistic approach is Algorithm \ref{alg:optimistic}. Chase steps are scheduled according to a suitable policy which maximizes performance; we discuss this further below. Each chase step's writes are allowed to proceed, and each write is checked to determine whether it changes the answer to a previously-posed read query by a lower-priority update. If so, the reader is aborted, together with any updates which themselves have read from the reader.

\begin{algorithm}
\caption{Optimistic scheduler template}
\label{alg:optimistic}
\begin{algorithmic}
\STATE choose next chase step \texttt{C} to schedule
\STATE \COMMENT{suppose \texttt{C} belongs to \texttt{u} which has number $j$}
\STATE execute \texttt{C}
\FORALL{writes \texttt{w} performed by the step}
\FORALL{stored read queries \texttt{q} of updates numbered $i > j$}
		\IF {\texttt{w} retroactively changes the result of \texttt{q}}
	  \STATE abort update number $i$ and any others who have read from it
		\ENDIF
\ENDFOR
\ENDFOR
\STATE \texttt{Q} := read queries actually performed by \texttt{C}
\STATE store \texttt{Q} for future checks
\end{algorithmic}
\end{algorithm}

In this algorithm, the scheduler's queue of updates that are ready to take a step is populated asynchronously, as updates complete frontier operations and return control to the scheduler. This means that chase steps may indeed be scheduled while another update's step is waiting for frontier operations. This may appear to violate our logical assumption that interleaving only occurs at chase step level. However, as we noted before, this is only a logical constraint, and it is not difficult to maintain. A chase step performs all its writes and violation read queries before asking for user input, so those cannot pose a problem. A step of a backward chase performs no further reads. A forward chase step may perform some correction queries after the user has performed a frontier operation; however, those queries can be logged by the scheduler and checked against any writes that occur properly after them. 

Algorithm \ref{alg:optimistic} enforces conflict serializability through a property which is strictly stronger than chase step safety; we call this property \emph{strong safety}:

\begin{definition}[Strong safety]
~Chase step $C$ is \emph{strongly safe} to perform for update $u$ with number $j$ on database $D$ iff for all $\tau' = \tau_1 \cdot \tau_2 \cdots \tau_{j-1}$, where each $\tau_i$ is a terminating extension for update number $i$ with respect to database $\tau \cdot \tau_1 \cdot \tau_2 \cdots \tau_{i-1}(D)$, we have that for all read queries $q$ that $u$ may perform in step $C$, and all prefixes $\tau''$ of $\tau'$,
\begin{equation*}
q(D) = q(\tau \cdot \tau''(D)).
\end{equation*}
\end{definition}

That is, a step is strongly safe to perform iff all the answers to its read queries have completely stabilized with respect to writes by lower-numbered updates. 

We briefly discuss the implementation of checking whether a write retroactively changes the result of a previously posed read query. Because read queries come in three very specific forms, this is not hard. Correction queries can be checked against writes without needing to access the database; a given tuple write changes the answer to a correction query either on all databases, or on none. For example, if a correction query asks for all tuples containing variable $x_2$, a write changes the answer iff the tuple written contains $x_2$. 

Violation queries are somewhat more complex, and here the check to be performed requires accessing the database. However, the process is still simple, as a write can only change the answer to a violation query in a limited number of ways. For example, an insert can do so in two ways. It can contribute to the creation of a join result among relations on the LHS of a mappings, so that a new witness appears. Alternatively, it can provide the last tuple that makes a tuple appear in the join of relations on the RHS of some rule. If the new RHS join tuple matches a previously existing witness, a violation is removed. Based on the type of the write (insert or delete) it is possible to perform the check by posing a single query which combines the original violation query with information about the new tuple. Tuple modifications are (conservatively) treated as a delete followed by an insert.

Algorithm \ref{alg:optimistic} is a template which can be instantiated in a variety of ways with respect to the scheduling policy used and how cascading aborts are determined.

\subsection{Determining cascading aborts}

When an update number $i$ aborts, all other updates which have read data affected by its writes must abort as well. As aborts are highly undesirable, it is important to develop accurate algorithms for determining such dependencies between updates.

A na\"ive, strawman algorithm (\texttt{NA\"IVE}) would abort all updates numbered $j > i$. This is sufficient to guarantee correctness, but is clearly not optimal. It is much better to keep track explicitly of dependencies between updates; whenever update $i$ issues a read query $q$, the system determines not just the answer to $q$, but also a list of updates with numbers $k < i$ whose writes directly influence the answer, that is, updates on which update $i$ now has a \emph{read dependency}.

\subsubsection{Our algorithms}
Computing read dependencies can be done in two different ways, one more precise and more expensive than the other. As before, correction queries are the easy case: given a logged list of all previous writes, it is easy to determine which ones change the result of a given query. If the list is kept in memory, the dependencies can be computed without querying the database. Violation queries, however, cannot be processed so simply without sacrificing precision.

The simpler of our two algorithms, \texttt{COARSE}, does not query the database to determine read dependencies caused by violation queries. When it processes a violation query that involves a set of relations $\{R_1, R_2, \cdots R_k\}$, it assumes that any update which has previously written any tuple to one of the $R_i$s may be the source of a read dependency. This is a conservative overestimate of the real read dependencies, so correctness is guaranteed. 

The second algorithm, \texttt{PRECISE}, trades off run time for precision. \texttt{PRECISE} determines accurately, for each violation query $q$, which previous writes have changed the answer to the query. That is, it finds all those updates that have performed some write such that the answer to $q$ would be different if the write had not yet been performed. As such, it comes very close to the theoretical optimum in detecting only true read dependencies. However, achieving this precision requires asking relatively complex queries on the database; the queries are closely related to those used to determine conflicts in Algorithm \ref{alg:optimistic}.

\subsubsection{Complexity}

\texttt{COARSE} has linear time complexity in the number of writes performed so far on the database by updates which may still be aborted. For \texttt{PRECISE}, the dominating contribution to the complexity is the cost of the joins in the queries; these joins are dictated by the mappings (as in Example \ref{ex:violationquery}). In the worst case, their time complexity is polynomial in the size of the database and exponential in the join arity, i.e. in the number of atoms per side of a mapping. However, since the join predicates are a function of the mappings and thus known up front, it is possible to improve performance by appropriate indexing and join implementation.

\subsection{Choosing a scheduling policy}

Ultimately, the performance of our concurrency control algorithm is measured in terms of throughput as well as the number and kind of aborts. The relative weighting of these two factors depends on the specific real-world setting that the Youtopia repository operates in, and we expect it to vary. We are developing a general and flexible set of scheduling policies that will be applicable to a wide range of parameters. Here, we mention some of the factors involved.

Ideally, update steps should be scheduled as soon as possible, and with as few aborts as possible. This means that the number of conflicts must be reduced as far as possible. Some form of static analysis of the updates with respect to their expected future reads and writes is likely to be helpful; however,  such analysis may not always yield meaningful results. It may be more useful for the scheduler to develop heuristics about expected operations, based on data such as previous update behavior in the system. Based on these heuristics, updates less likely to conflict can be allowed to interleave more aggressively and/or run in parallel.

Once an update is chosen to run, the scheduler has to decide how long it may run. In some cases, it may be beneficial to allow updates to complete an entire deterministic stratum; in others, strict step-by-step scheduling is best.

When an update blocks awaiting a frontier operation, the scheduler may permit other updates' steps to proceed, as explained earlier. The decision will depend on the expected costs and benefits; for example, if the frontier operations involve a table that has a good track record in terms of fast user response, the scheduler may choose to block in anticipation of the frontier operation. This is particularly true if the blocking update is ``worth waiting for" -- because, for example, many of the other running updates are likely to read from the relations involved in its frontier tuples.

\section{Experiments}

Our experiments compare the performance of the algorithms \texttt{NA\"IVE}, \texttt{COARSE} and \texttt{PRECISE} with respect to the number of cascading aborts and execution time. Since Youtopia's paradigm of relational CDI is new and there are no real-world datasets for us to benchmark our algorithms against, we have used synthetic data and mappings, as explained below. Further, we needed to find a way to simulate frontier operations. Our code does this by choosing an option uniformly at random among all available alternatives for a frontier operation. In practice, this has the additional advantage of making all chases terminate, even when mappings have cycles: a unification (rather than expansion) operation is chosen sooner or later on every forward chase path.

Our experiments are run on a database of 100 relations, each randomly generated to have between one and six attributes. The relations are connected by mappings; each mapping is created by choosing a random subset of one to three relations for the LHS and another for the RHS. Smaller sets have higher probability, as humans are highly unlikely to create mappings with more than one or two atoms on either side. The remaining step in mapping generation is the choice of variables in the atoms; this is done randomly, with care taken to ensure that the mappings contain inter-atom joins as well as constants. Any constants used come from a small (size 50) fixed set of random strings.

Generating the initial database is performed using our update exchange techniques themselves, with simulated user interaction; it is not easy to obtain an interesting database that satisfies an arbitrary, potentially cyclic, set of tgds using another method. We generate ten thousand initial tuples. The relations receiving those tuples are chosen uniformly at random, and the attribute values come from the same set of constants that was used in mapping generation. By keeping the domain of attribute values small, we ensure that joins between relations are highly likely to be nonempty, and that mappings are consequently highly likely to fire if these tuples are inserted. We insert these initial tuples into the database; each insertion sets off a forward chase which only ends when all constraints are satisfied.

We test our algorithms in several settings which vary in the number of mappings. We vary this number from 20 (a sparse setting) to 100 (a dense one). Settings with denser mappings are likely to exhibit longer chase runs, more writes, and therefore more conflicts and aborts; indeed, this is borne out by our results. In our runs, the set of mappings we used is monotonically increasing -- that is, our experiments with 40 mappings involve the mappings used for the run with 20 mappings as well as 20 others, and so on. In all cases, the initial database is the same and satisfies all 100 mappings.

We show results on two workloads, each of 500 updates. The first consists entirely of inserts, the second of eighty percent inserts and twenty percent deletes. Each update in each workload is started by an insert or delete operation generated randomly and independently. First, the receiving relation is chosen uniformly at random. In the case of inserts, the values in the inserted tuples are chosen with equal probability to be fresh or from the previously mentioned set of constants. In the case of deletes, the tuple to delete is chosen uniformly at random from the relation. In the mixed insert/delete workload, the order of the updates is then randomized to ensure that the runs do not involve alternating large batches of inserts and deletes. The scheduling algorithm used in all our experiments is a round-robin policy that interleaves chases at the level of individual steps.  All runs are allowed to run to termination, and each data point is obtained as the average of 100 runs.

\begin{figure*}

\begin{minipage}[b]{0.66\columnwidth}
\centering
\includegraphics[width=\linewidth]{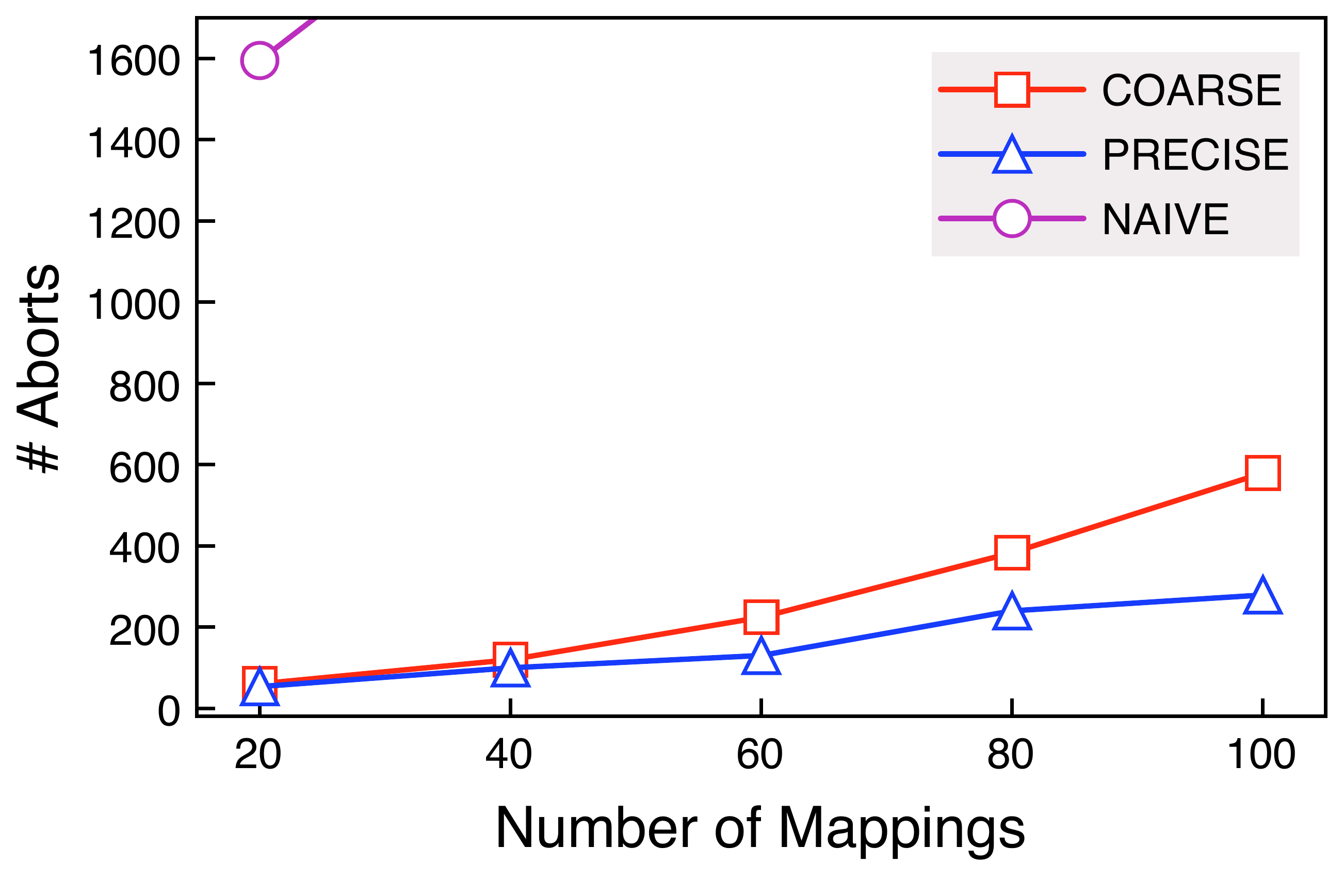}
\end{minipage}
\begin{minipage}[b]{0.66\columnwidth}
\centering
\includegraphics[width=\linewidth]{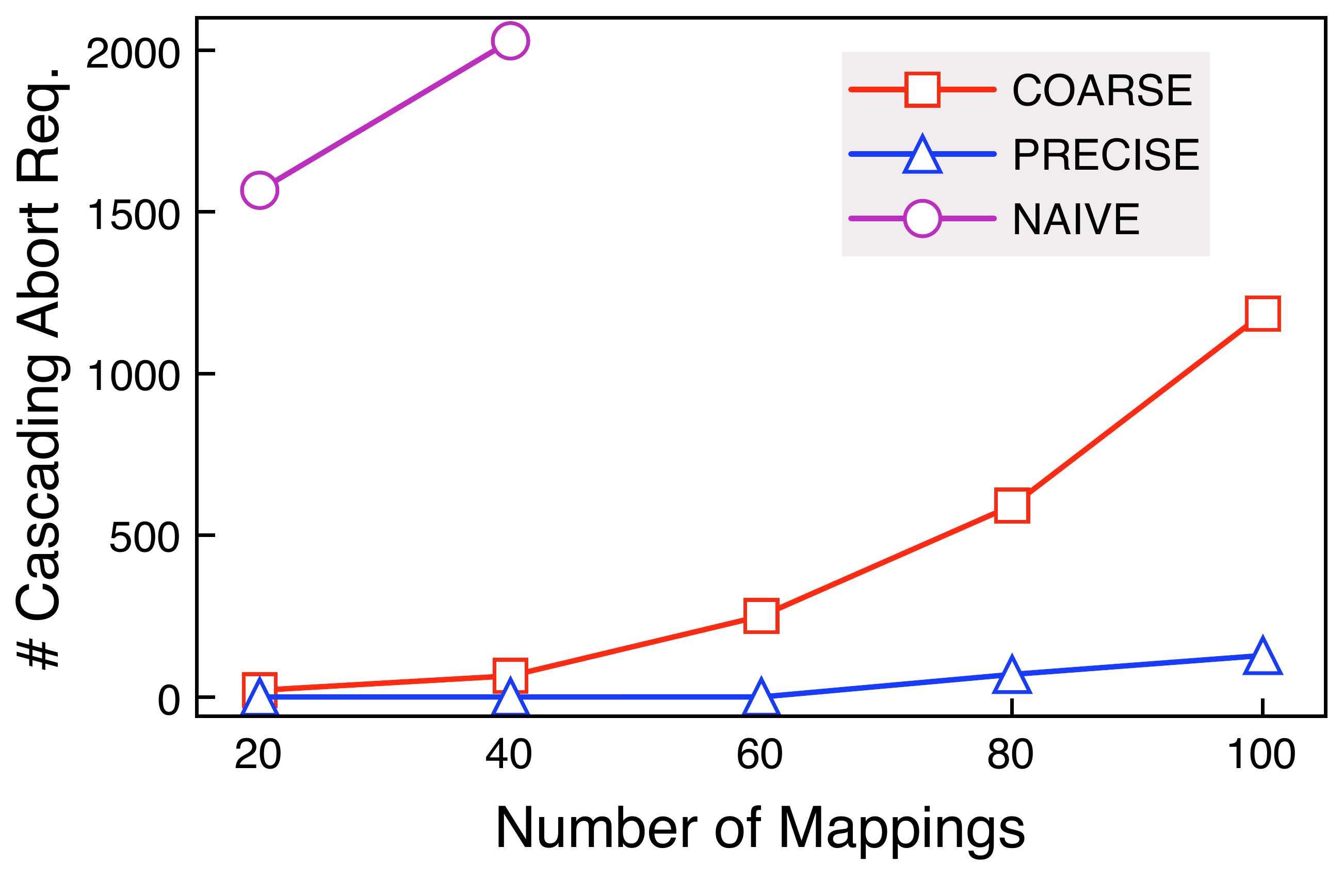}
\end{minipage}
\begin{minipage}[b]{0.66\columnwidth}
\centering
\includegraphics[width=\linewidth]{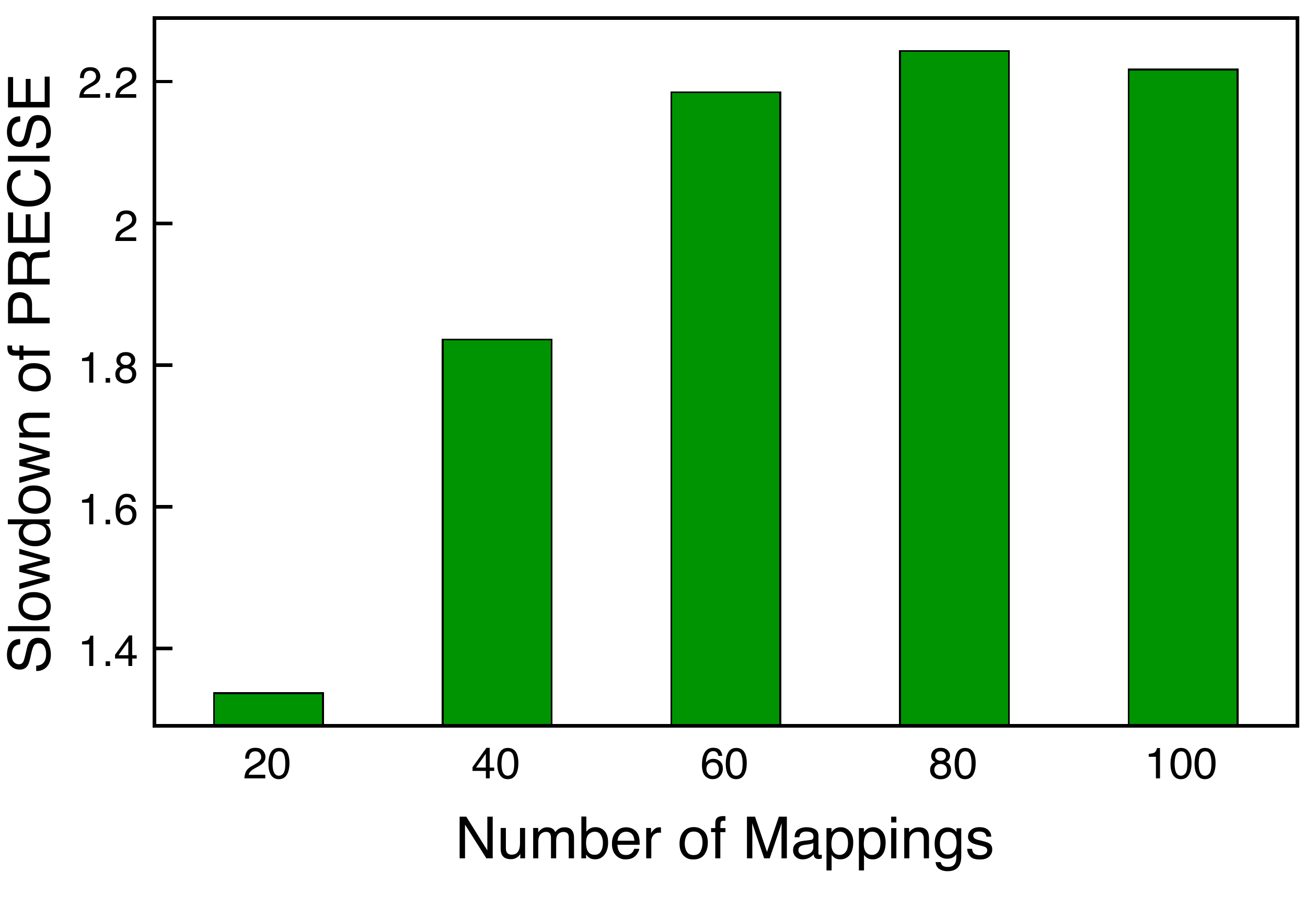}
\end{minipage}
\caption{Results for the all-insert workload}\label{fig:insertexp}
\end{figure*}

\begin{figure*}

\begin{minipage}[b]{0.66\columnwidth}
\centering
\includegraphics[width=\linewidth]{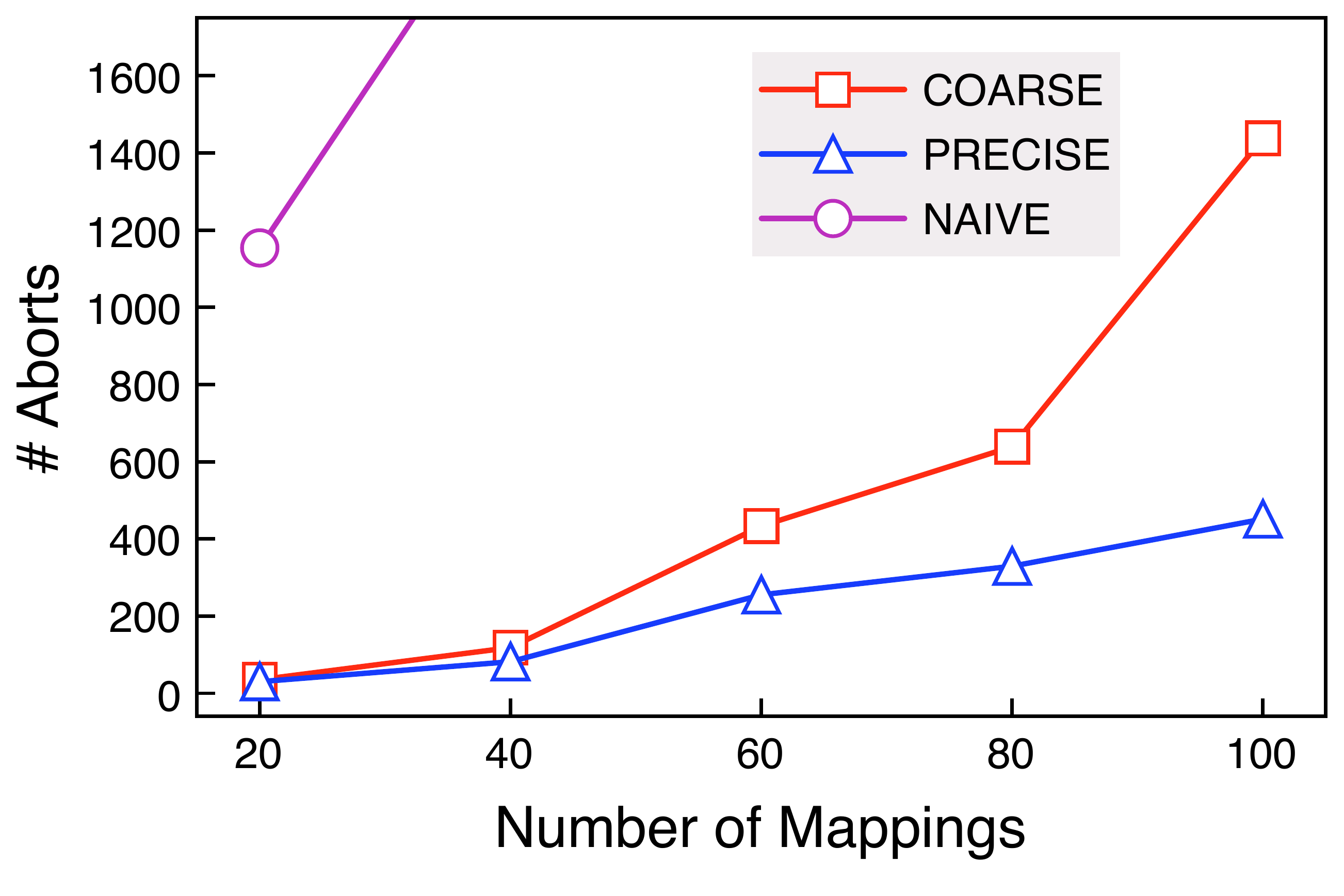}
\end{minipage}
\begin{minipage}[b]{0.66\columnwidth}
\centering
\includegraphics[width=\linewidth]{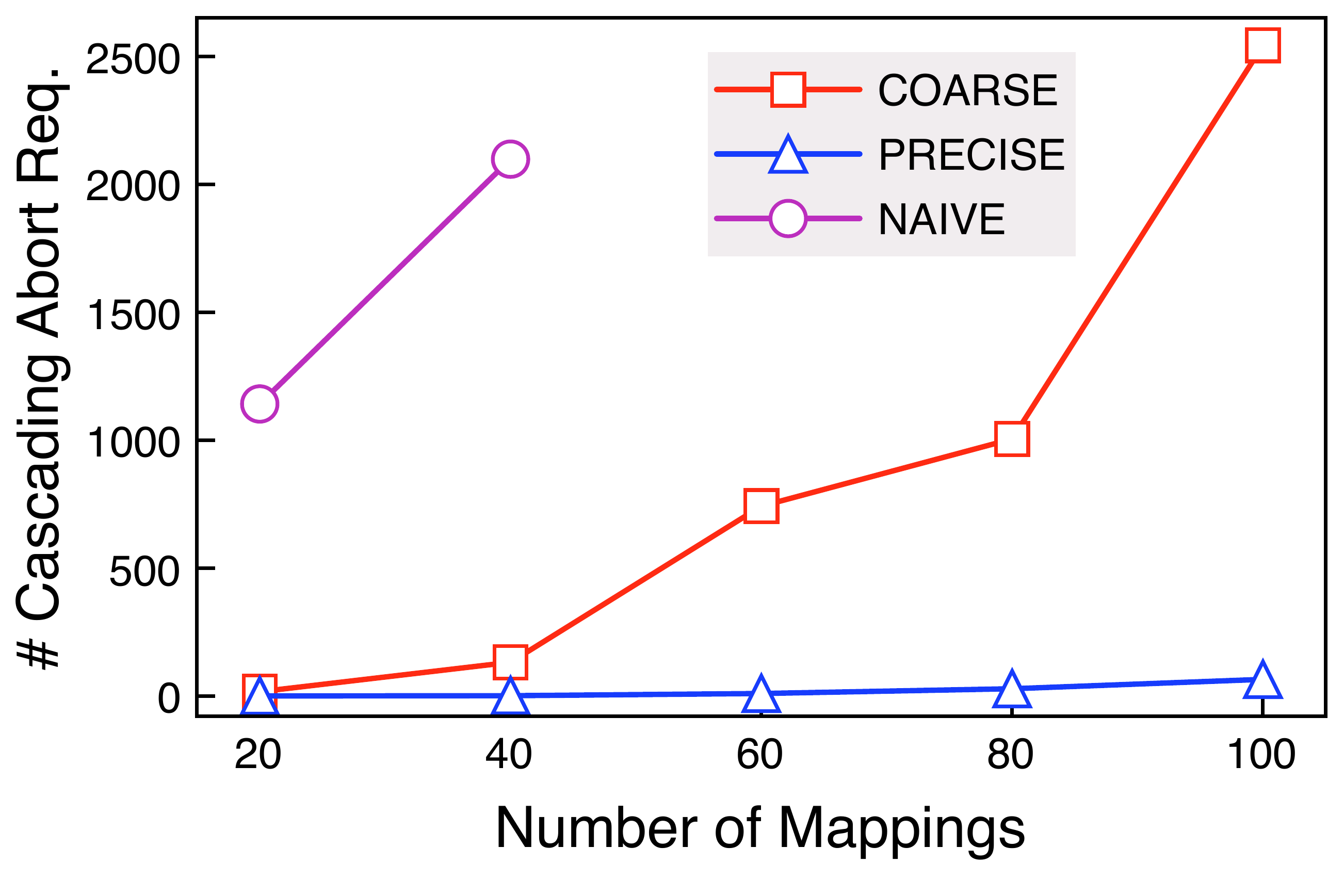}
\end{minipage}
\begin{minipage}[b]{0.66\columnwidth}
\centering
\includegraphics[width=\linewidth]{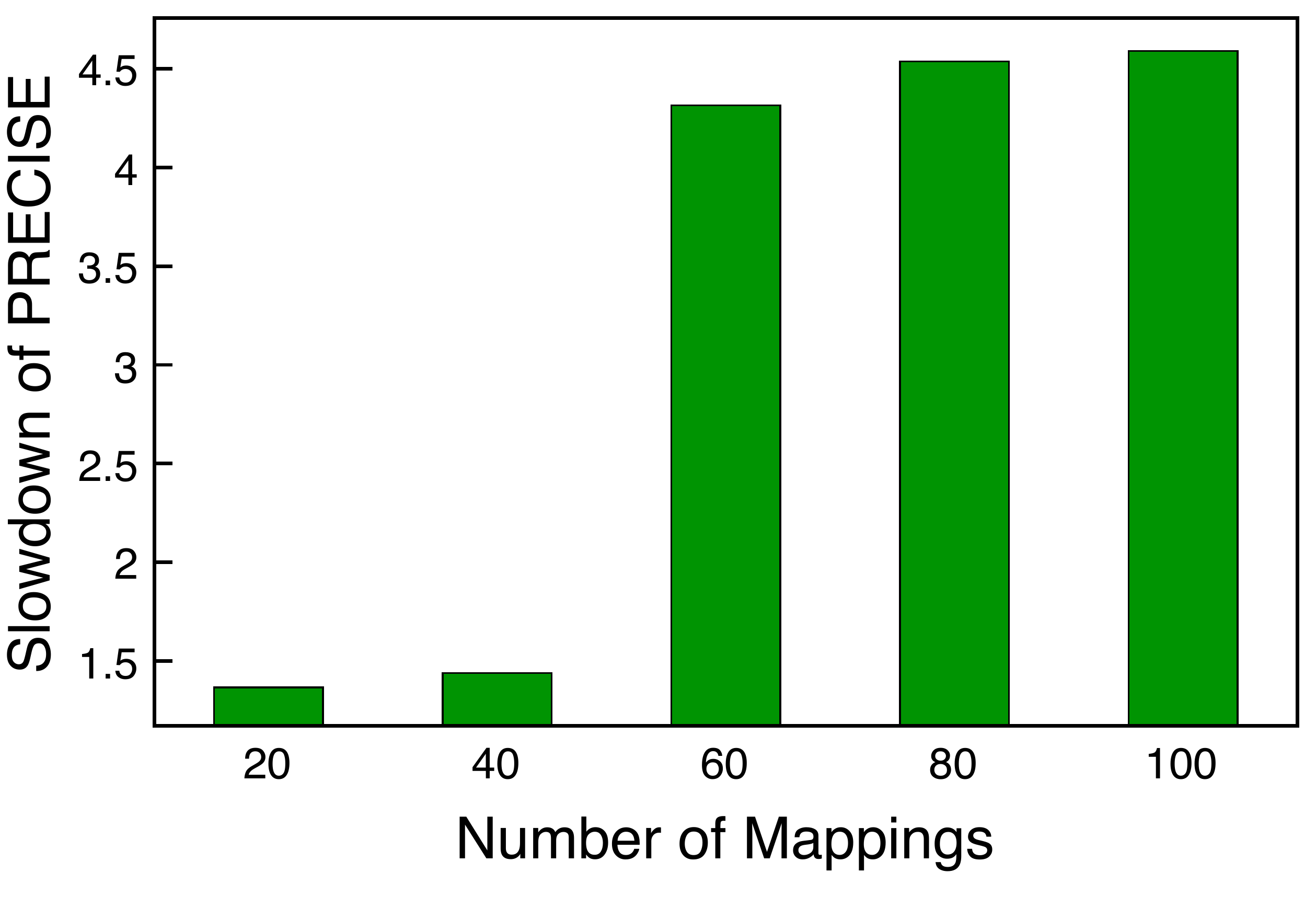}
\end{minipage}
\caption{Results for the mixed workload}\label{fig:mixedexp}
\end{figure*}

Our results are shown in Figures \ref{fig:insertexp} and \ref{fig:mixedexp}. The first graph of each figure shows the total number of aborts encountered during the run. Clearly, both \texttt{COARSE} and \texttt{PRECISE} outperform \texttt{NA\"IVE} significantly. We only show the first few points for \texttt{NA\"IVE}, as the huge performance difference is apparent even with very sparse mappings.

The second graph shows the total number of \emph{cascading abort requests} during each run. This is the number of times during the run that the algorithm requests an abort even though the update involved is not in direct conflict with a just-performed write. Thus, this is the number of \emph{purely cascading} aborts requested. It does not have a direct correspondence to the total number of aborts observed during the run. This is because aborts are not performed as soon as they are made necessary by a write, but only once control is returned to the scheduler. In the meantime, abort information related to various writes performed by a chase step is collected and collated. Updates are frequently marked for abortion multiple times during that phase; however, the scheduler only performs aborts based on the consolidated information. This metric clearly shows the difference between \texttt{COARSE} and \texttt{PRECISE}; indeed, in scenarios with lower mapping density, \texttt{PRECISE} requests \emph{no} cascading aborts.

The final figure shows the relative time penalty associated with the use of \texttt{PRECISE} over \texttt{COARSE}. This is the ratio of per-update execution times for each algorithm. The per-update execution time is obtained by dividing the total time for the run by the number of updates which actually ran (i.e., the original 500 plus the number of aborts). In this way, we adjust for the fact that runs with \texttt{PRECISE} involve a lower number of total update executions. The graph shows the relative slowdown rather than the per-update execution times themselves. The actual execution time increases for each algorithm with the number of mappings, which is not surprising, since more mappings require more read queries and more writes to be performed. Our purpose here is orthogonal to demonstrating this increase: we aim to show the overhead of using \texttt{PRECISE} instead of \texttt{COARSE} across a variety of mapping density settings.

Our experiments show that \texttt{COARSE} and \texttt{PRECISE} significantly alleviate the cascading abort problem. As expected, \texttt{PRECISE} does best, but at the cost of an increase in execution time. In practice, we expect that the reduction in the number of aborts will be so important to Youtopia users that the increased execution time of \texttt{PRECISE} will be acceptable. However, if this time overhead should prove too large, it is also possible to use a hybrid policy combining \texttt{COARSE} and \texttt{PRECISE} on a per-update basis. An update which is particularly important and which should not be aborted spuriously -- perhaps because it has already aborted several times -- can have its read dependencies determined using \texttt{PRECISE}, so that it only aborts when it absolutely needs to. For less important updates, \texttt{COARSE} can be used.

We also remark that the absolute number of aborts across all our experiments remains quite high; this underscores the need for a good scheduling policy to minimize the number of aborts that are non-cascading, i.e., due to genuine conflicts.

\section{Related work in CDI} \label{sec:relatedDataIntegration}

Sections 2 through 5 of the paper include specific references to related work where appropriate. Here, we take a broader perspective and give a brief overview of existing solutions for CDI tasks.

There is a growing body of work which adapts classical data integration ideas to the community setting, including substantial theoretical work \cite{FuxmanKMT06:Peer, CalvaneseGLR04:Logical, FranconiKLS03:Robust}. Systems like Orchestra \cite{TaylorI06:Reconciling}, Piazza \cite{HalevyIST05:Schema}, Hyperion \cite{Gianolli05:Data} and the system introduced in \cite{KatsisDP08:Interactive} focus on maintaining data utility despite significant disagreement. However, none of these enable best-effort cooperation to its fullest extent. They all come with some centralized logical component that is an extensibility bottleneck; usually this is either a global schema or an acyclicity restriction on the mappings, or both. In addition, they do not provide facilities for users to manage the metadata collaboratively. 

Best-effort cooperation is fundamental in projects such as Cimpl \cite{DoanRCDLMSS06:Community} and MOBS \cite{MccannKSVSD05:Integrating}. However, these are designed for settings where disagreement is relatively mild and there is a meaningful notion of an authoritative version on the data and metadata; not all CDI scenarios fall into this category.

Wikipedia and related systems have been higly successful at addressing all three of the CDI aspects; they are highly cooperative, have a robust disagreement handling infrastructure, and achieve great success because of the utility of the data they contain. However, they only work for unstructured or mildly-structured data. Furthermore, Wikipedia does not fully support integration, requiring most data cleaning and management tasks to be done by hand.

Google Base is a very interesting point in the CDI design space. It allows best-effort cooperation by making it easy for anyone to add data, whatever its content or format; further, it does not require any agreement on the data. This raises a potential utility problem -- how can one easily query the database if there is no global schema? This is solved by forcing na\"ive users to interact with the database through predefined views such as ``Hotels'', ``Recipes'', etc. Nevertheless, Google Base does not perform full data integration, and there is no support for collaborative data management. On the other hand, scientific data sharing portals such as BIRN \cite{Birn} and GEON \cite{Geon} allow significant cooperation, perform substantial integration and permit some disagreement, but are not lightweight general-purpose solutions that are as easy for nontechnical users to work with as Google Base.

Given the lack of an ideal CDI solution, real-world Web communities often just use a shared database. This happens, for example, on many vertical social networking sites. Even when the community members themselves are not technically savvy, the data they share can have fairly substantial structure, as on the  craft site Ravelry \cite{Ravelry}. Shared databases provide good data utility, but cannot handle disagreement or nontrivial cooperation. Schema extensibility in particular is a real problem, even among nontechnical users. For example, recent months have seen a debate on Ravelry about the site's (current) inability to meet users' wishes for tables devoted to additional crafts \cite{ForbesRav}.

CDI and the Youtopia system are highly compatible with the Dataspaces vision \cite{HalevyFM06:Principles}. Indeed, a Youtopia repository can be seen as a dataspace. However, our initial focus is more restricted: we set out to enable relational data sharing among members of a relatively knowledgeable and motivated community. We believe this setting is associated with unique challenges and opportunities, and deserves a dedicated solution. Such a solution could profitably be integrated into any other dataspace designed for a setting where highly structured data is shared.

\section{Future work}
\label{sec:futureWork}

Throughout the paper, we have mentioned possible extensions to our models and algorithm, as well as areas in which we have ongoing work already. We also plan to add support for arbitrary tuple modifications rather than just global null-replacements; those pose new challenges as they can cause both LHS-violations and RHS-violations, thus setting off a forward and a backward chase simultaneously. 

There are many interesting foundational questions related to our new chase model. For example, can a database always be brought to a state where it satisfies all mappings using only a finite sequence of frontier operations?

Last but not least, we are developing user interfaces that facilitate frontier operations by presenting meaningful information about the provenance of frontier tuples.

\bibliographystyle{abbrv}

\end{document}